# 10032
# Proposal of Automatic FPGA Offloading for Applications Loop Statements

Yoji Yamato[a],*

[a] NTT Network Service Systems Laboratories, NTT Corporation, Japan
E-mail address: yoji.yamato.wa@hco.ntt.co.jp

**Abstract**

In recent years, with the prediction of Moore's law slowing down, utilization of hardware other than CPU such as FPGA which is energy effective is increasing. However, when using heterogeneous hardware other than CPUs, barriers of technical skills such as OpenCL are high. Based on that, I have proposed environment adaptive software that enables automatic conversion, configuration, and high-performance operation of once written code, according to the hardware to be placed. Partly of the offloading to the GPU was automated previously. In this paper, I propose and evaluate an automatic extraction method of appropriate offload target loop statements of source code as the first step of offloading to FPGA. I evaluate the effectiveness of the proposed method using existing applications.

## 1. Introduction

Recently, it is said that Moore's Law will slow down and CPU's semiconductor density cannot be expected to double in 1.5 years. Based on this background, applications with hardware such as FPGA (Field Programmable Gate Array) or GPU (Graphics Processing Unit) are increased. In particular, FPGA has better power efficiency than CPU, and it is expected to reduce power consumption by using FPGA. For example, Microsoft's search engine Bing tries to use FPGA [1] to improve search effectivity and Data Center energy consumption. Amazon AWS (Amazon Web Services) [2] provides FPGA and GPU instances using Cloud technologies (e.g., [3]-[13]).

However, to obtain high performances using non-CPU hardware appropriately, programmers need to program and configure considering hardware specification and need to use expert skills such as OpenCL (Open Computing Language) [14] and CUDA (Compute Unified Device Architecture) [15]. These barriers are high for many programmers.

Along with the progress of IoT (Internet of Things) technology (e.g., Industrie 4.0 and so on [16]-[20]), IoT devices are increasing rapidly, and many IoT applications are developed using service coordination technologies such as [21]-[30].

In summary of backgrounds, systems with heterogeneous hardware such as FPGA, GPU and many IoT devices are expected to increase more and more, however barriers are high to utilize heterogeneous hardware effectively. In order to remove these barriers and utilize heterogeneous hardware easily and effectively, the platform that developers only write logics to be processed, then software will adapt to the deployed environments with heterogeneous hardware by converting, configuring automatically is expected, I think.

Because Java [31] is insufficient for environment adaptation with performances, I have proposed environment adaptive software which execute once written applications with high performance by automatic code conversion and configurations so that FPGA, GPU, IoT devices or others can be used on deployed environments appropriately. As elementary technology of environment adaptive software, I have also achieved automatic GPU offloading of some applications software [32][33]. In this paper, I propose a method for automatic offloading of appropriate loop statements of application software as the first step of offloading to FPGA, which is energy effective compared to CPU. I implement the proposed method and evaluate the effectiveness of FPGA offloading using existing applications.

## 2. Existing Technologies

To control heterogeneous hardware such as FPGAs, many core CPUs and GPUs uniformly, OpenCL specification are appeared and its SDK (e.g., [34][35]) are spread. For GPGPU (General Purpose GPU) that uses GPU parallel computation power not only for graphics processing (e.g., [36]) but also for other purposes, CUDA is a major environment. OpenCL and CUDA need not only C language extension grammars but also additional hardware oriented descriptions such as memory copy between FPGA devices and CPUs. Because of these difficulties, there are few OpenCL or CUDA programmers.

Comparing to OpenCL, for easy using of heterogeneous hardware, there are technologies that specify parallel processing areas by specified directives and compilers convert these codes into device oriented codes based on specified directive meanings. In terms of directive-based specifications, OpenACC [37] is one example, and in terms of directive-based compilers, PGI compiler [38] is one example. For example, users specify OpenACC directive "#pragma acc kernels" on Fortran/C/C++ codes to process them in parallel, and the PGI compiler checks the parallel processing possibility, outputs and deploys binary files to execute on CPUs and GPUs. For Java,

IBM JDK supports GPU offloading based on Java lambda expression [39].

In this way, OpenCL, CUDA, OpenACC and other technologies enables FPGA or GPU offload processing itself. However, although processing on FPGA or GPU can be performed, high performance is difficult to achieve. For example, there are automatic parallelization technologies like the Intel compiler [40] for many core CPU. Those extract possible areas of parallel processing such as for and while loop statements. However, naive parallel processing performances with FPGAs or GPUs are not high because of overheads of CPU and FPGA/GPU devices memory data transfer. To achieve high performances with FPGA/GPU, high skill programmers need to tune using OpenCL/CUDA or appropriate offloading areas

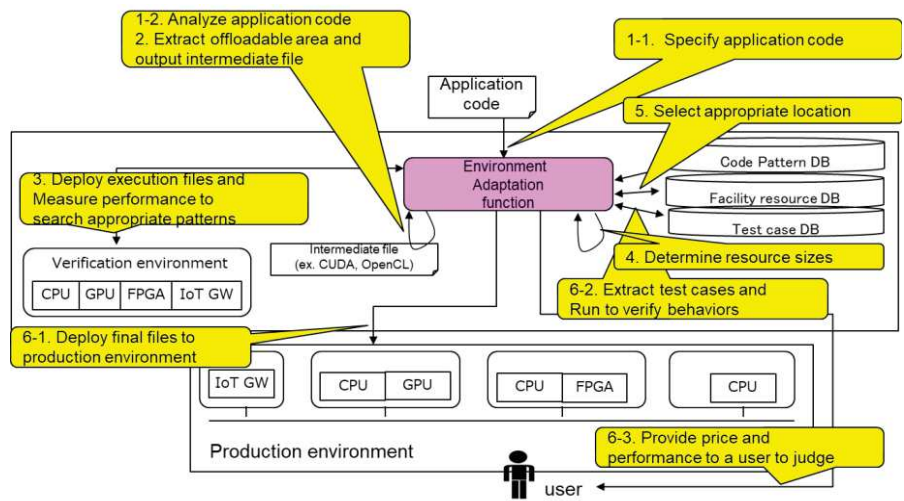

Fig. 1: Flow of environment adaptive software need to be searched for.

## 3. Proposal of Automatic FPGA Offloading for Loop Statements

### 3.1 Flow of environment adaptive software

To achieve software adaptation to the environment, I have proposed the following processing flow of environment adaptive software (Figure 1). The environment adaptive software is executed in cooperation with functions including an environment adaptation function which is a main function, a verification environment, a production environment, a test case DB (using Jenkins, etc [41][42]), a code pattern DB, a facility resource DB.

### 3.2 Consideration points for automatic FPGA offloading

For code analysis in Step 1, parse and analyzation of application code is executed using a parsing tool such as Clang [43]. Code analysis is hard to generalize because it is necessary to analyze considering devices to be offloaded. However, in analyzing step, it is necessary to grasp the structure of the source code such as loop statements, reference relations with the variables and grasp function blocks of specified processing or calling a specified function library such as FFT processing in common. It is very difficult for machine to detect function blocks automatically, I will use similarity detection tools such as Deckard to judge code similarity. In addition, Clang is a tool for C/C++, it is needed to choose a tool suited to the language to be parsed.

In Step 2-3, it is necessary to consider processing according to the offload destination, such as FPGA, GPU and IoT GW, it is assumed that processing functions are plugged in for each offloading destination. In general, regarding to performances, it is difficult to automatically detect the configuration that will be the maximum performance at one time, thus we repeatedly try the offload patterns in the verification environment several times to detect an appropriate offload pattern by an evolutionary computation method. Regarding to GPU, I achieved automatic offloading for some applications in the previous work [32]. Therefore, this paper studies FPGA offloading of application software.

Applications that users want to offload are various. However, typical processing that require a lot of computation time are many number of loops such as image analysis ([44][45]) for movie processing, machine learning processing for analyzing sensor data, or so on. Therefore, in this paper, first target to FPGA offloading is also loop statements.

For GPU offloading, OpenACC only specify #pragma acc kernels directive so that specified loop statements can be executed on GPU or CUDA can describe more detail control. To control FPGA, OpenCL can describe detail control like CUDA and High Level Synthesis (HLS) tools can specify more abstract control like OpenACC. Regarding to OpenCL, 10 steps descriptions are needed. "Prepare devices, Prepare kernels, Allocate devices memory, Transfer data from hosts to devices, Configure variables of kernel functions, Execute kernel functions, Transfer data from devices to hosts, Release devices memory, Release kernels, Release other objects such as devices".

To extract appropriate offloading areas from general CPU programs automatically, my previous work of [32] firstly checks all loop statements to determine whether they can be processed or not and secondly executes performance verifications repeatedly in the verification environment using Genetic Algorithm (GA) [46] for processable loop statements to search for the appropriate offload pattern. However, code compiling to FPGA takes several hours in general, and performance measurements of many patterns like [32] are difficult. Therefore, it is assumed that the number of performance measurement will be reduced after narrowing down the pattern for performance measurement with actual FPGA.

## 3.3 Automatic FPGA offloading method

The method firstly parses source codes to be offloaded. It understands the loop statements and variables information according to the language of the source codes.

Next, a process to narrow down candidates is performed for whether or not to try FPGA offloading for the loop statements. Arithmetic intensity can be a one indicator of whether a loop statement has an offload effect. Arithmetic intensity is an index that increases when the number of loops and the amount of data are large, and decreases when the number of accesses is large. Processing with high arithmetic intensity is a heavy processing for the processor and takes time. Therefore, an arithmetic intensity analysis tool analyzes the arithmetic intensity of the loop statement and narrows down the high intensity loop statements for offloading candidates.

Even if it is a high intensity loop statement, it is a problem that it consumes FPGA resources excessively when it is processed on FPGA. Therefore, we move on to estimating the amount of resources when processing high intensity loop statements on FPGA. When compiling to FPGA, it is converted from a high level language such as OpenCL to a hardware level language such as HDL, and actual wiring processing is performed based on hardware level language. At this time, wiring processing and so on take a lot of time, but it takes only a minute until to extract HDL as the intermediate state. Since resources such as Flip Flop and Look Up Table used in FPGA can be estimated at the HDL level, the amount of resources used can be known in a short time even if compiling is not completed. Since the target loop statement is converted into a high level language such as OpenCL and the resource amount is calculated from the OpenCL, the arithmetic intensity and resource amount when the loop is offloaded are determined. In this method, the loop statements with high resource efficiency are further narrowed down as offload candidates. High resource efficiency means (arithmetic intensity/resource amount) is high.

Here, two processes are required to make a loop statement into a high level language such as OpenCL. One is to divide a CPU processing program into a kernel (FPGA) program and a host (CPU) program based on the syntax of a high level language. The other is to include techniques for speeding up for loop statements. In general, there are techniques for speeding up using FPGA such as local memory cache, stream processing, multiple instantiation, loop statement expansion, integration of nested loop statements, memory interleaving and so on. Depending on the loop statement, these may not have an absolute effect, but are often used as methods for speeding up.

Next, because some loop statements with high resource efficiency are selected, plural offloading patterns that measure the performances using these loop statements are generated. There are types of speed up in FPGA, one type of processing speeds up by concentrating the amount of FPGA resources to one process, and the other is the speedup by distributing FPGA resources to multiple processes. The method generates patterns for the selected single-loop statements, compiles them to run on FPGA, and measures the performances. Then, the method generates combination patterns for the single loop statements that can be accelerated,

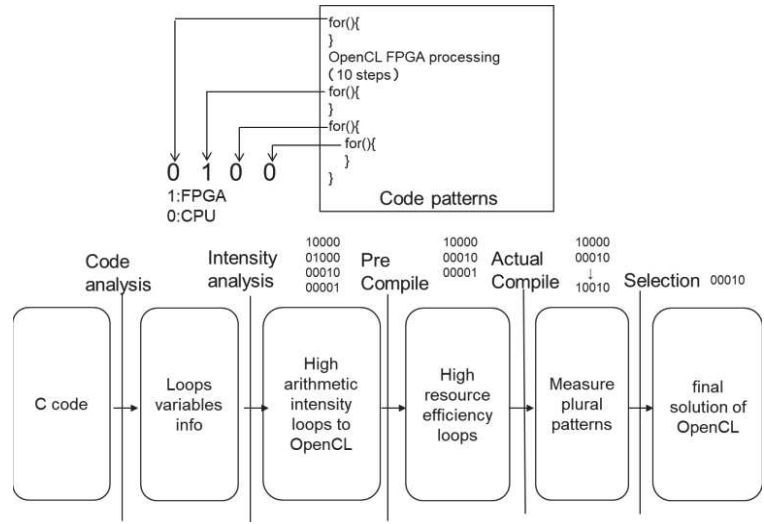

Fig. 2: FPGA offloading method of loop statements

and measures the performance in the same way.

Finally, a high-speed pattern is selected as the solution among the multiple patterns whose performance has been measured in the verification environment.

In this way, the method focuses on loop statements with high calculation density and resource efficiency, creates offloading patterns, and searches for patterns at high speed through actual measurements in the verification environment (Figure 2).

## 4. Implementation

In this section, I explain the implementation of the proposed method. To confirm the method effectiveness, we use C/C++ language applications for offloading applications and Intel PAC with Intel Arria10 GX FPGA for FPGA.

To control FPGA, we use Intel Acceleration Stack Version 1.2 (Intel FPGA SDK for OpenCL 17.1.1，Quartus Prime Version 17.1.1).

Here, I explain the outline of the implementation. We implement the implementation by Python 2.7.

When a C/C++ application is specified, the implementation analyzes C/C++ code, detects "for" loop statements and understand the program structure such as variables data used in the "for" statements. For parsing, our python program uses parsing libraries of LLVM/Clang 6.0 libClang python binding.

Next, the implementation executes the arithmetic intensity analysis tool to judge the possibility of the FPGA offloading effect of each loop statement, and obtains an index of arithmetic intensity determined by the number of loops, data size, number of accesses and so on. Only the number of top A loop statements with the highest arithmetic intensity are targeted. PGI compiler 19.4 is used for arithmetic intensity analysis. Although the PGI compiler is a GPU compiler, it can be used for arithmetic intensity analysis, so only the arithmetic intensity analysis part is used. To count loop number, we also can use gcov [47] or gprof.

The implementation then generates FPGA offloading OpenCL code for each loop statement with high arithmetic intensity. The OpenCL code is divided into the loop statement as an FPGA kernel and the rest as a CPU host program. When the FPGA kernel code is generated, the loop sentence is expanded by number B as a speed-up technique. The loop statement expansion process increases the amount of resources, but is effective for speeding up. The number of expansions is limited to a fixed number B, and the amount of resources does not become enormous.

The implementation then pre-compiles the A OpenCL codes using the Intel FPGA SDK for OpenCL, and calculates the amount of resources such as Flip Flop and Look Up Table to be used. The amount of resources used is displayed as a percentage of the total resource amount. Here, the resource efficiency of each loop statement is calculated from the arithmetic intensity and resource amount. For example, a loop statement with an arithmetic intensity of 10 and a resource amount of 0.5 has a resource efficiency of 10/0.5=20, a loop statement with a arithmetic intensity of 3 and a resource amount of 0.3 has a resource efficiency of 3/0.3=10, and the former is high. In loop statements, the implementation selects C OpenCL codes with high resource efficiency.

Next, the implementation generated patterns to be measured using selected C loop statements as candidates. For example, if the first, third and fifth loops are highly resource efficiency, the implementation generates and compiles an OpenCL patterns with #1 offloaded, #3 offloaded, and #5 offloaded. In the first measurement, the implementation generates patterns within D and conducts performance measurements on a server with FPGA in the verification environment. For performance measurement, the sample processing specified by the application to be accelerated is performed. For example, in the case of an application of Fourier transform, the performance is measured using the transform processing with sample data as a benchmark. Among them, if #1 and #3 offloading can be accelerated, the implementation generates a pattern with both #1 and #3 offloaded in the second measurement. Note that when generating a combination of single loop, the amount of resources is also a combination, so if it does not fit within the upper limit, the combination pattern is not generated.

The implementation finally selects the maximum performance pattern from plural measured patterns.

# 5. Evaluation

## 5.1 Evaluation method

### 5.1.1 Evaluated applications

I evaluate signal processing of time domain finite impulse response filter.

Time domain finite impulse response filter is a type of filter that performs processing in a finite time on the output when an impulse function is input to the system. There are various implementations. Among them, I use [48]'s C code and also use sample tests with it for performance measurement. When considering applications that transfer signal data from IoT devices over the network, to reduce network costs, it is assumed that signal processing such as filters are conducted devices sides. For this, the automatic FPGA offloading of signal processing has a wide range of applications, I think.

MRI-Q [49] processes MRI (Magnetic Resonance Imaging) 3D images. In an IoT environment, image processing is often necessary for automatic monitoring from camera videos and performance enhancements are requested in many cases.

### 5.1.2 Experiment conditions

For automatic FPGA offloading, we do not conduct many performance measurements like previous automatic GPU offloading study [32]. In the verification, the performance measurement results of the sample tests with multiple offload patterns in the verification environment are recorded together with the intermediate information such as arithmetic intensity, resource efficiency, HDL related information during compilation and so on. Through performance measurement, the highest performance pattern is the search solution, and the performance of the solution is evaluated as compared to performances of all CPU processing.

Conditions of experiments are as follows.

Offloading target number: Number of loop statements. 36 for time domain finite impulse response filter. 16 for MRI-Q.

Narrow down of arithmetic intensity: Narrow down to the top five loop statements of arithmetic intensity.

Number of loop statement expansions: 1. Expansion processing and multiple instantiations can often be accelerated as the amount of resources is used. Thus, in this verification, I confirm the effect of FPGA offloading with OpenCL without expansions.

Narrow down of resource efficiency: Narrow down to the top three loop statements in resource efficiency analysis. The implementation selects top three loop statements with high (arithmetic intensity / resource amount).

Number of measured offload patterns: 4. First time, the top three loop statement offload patterns were measured, and the second time was measured with the combination pattern of two loop statement offloads that were high performance at the first time.

### 5.1.3 Experiment environment

I use physical machines with Intel PAC with Intel Arria10 GX FPGA for FPGA offloading verifications. I use Intel Acceleration Stack Version 1.2 for FPGA control. Figure 3 shows an experiment environment and environment specifications. Here, a client note PC specifies C/C++ application codes, codes are tuned with try and error on a verification machine, and final codes are deployed in a running environment for users after tuning.

### 5.2 Performance results

As an application where manual acceleration is often performed on FPGA, we confirmed automatic accelerations of time domain finite impulse response filter.

Figure 4 shows the measurement results of how much performance has been improved by the proposed method implementation of automatic FPGA offloading. Figure 4 shows how many times the performance of the final solution is higher than the performances of all CPU processing. From Figure 4, it can be seen that the proposed method achieves 4.0 times performance for the time domain finite impulse response filter. It can be also seen that the proposed method achieves 7.1 times performance for the MRI-Q. As for the automation time, it takes about half day to automatically verifications of 4 patterns because it takes about 3

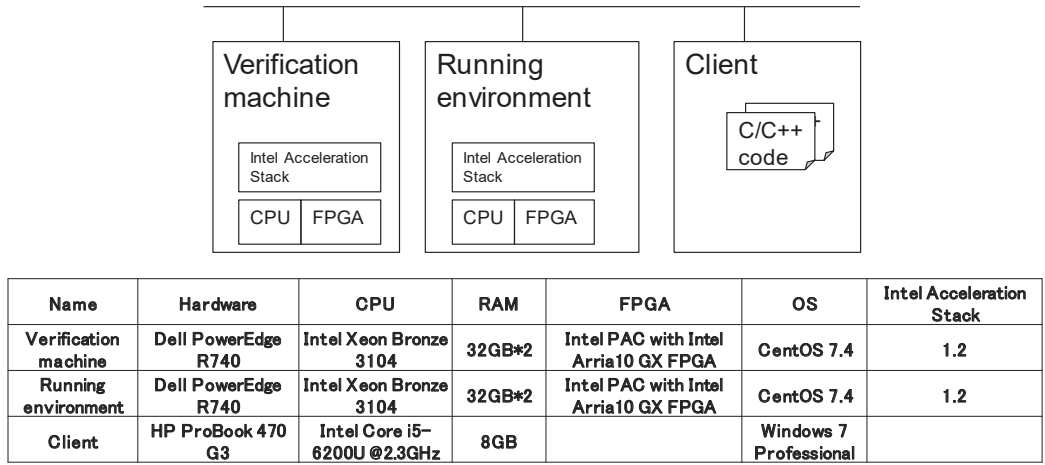

| Name | Hardware | CPU | RAM | FPGA | OS | Intel Acceleration Stack |
|---|---|---|---|---|---|---|
| Verification machine | Dell PowerEdge R740 | Intel Xeon Bronze 3104 | 32GB*2 | Intel PAC with Intel Arria10 GX FPGA | CentOS 7.4 | 1.2 |
| Running environment | Dell PowerEdge R740 | Intel Xeon Bronze 3104 | 32GB*2 | Intel PAC with Intel Arria10 GX FPGA | CentOS 7.4 | 1.2 |
| Client | HP ProBook 470 G3 | Intel Core i5-6200U @2.3GHz | 8GB | | Windows 7 Professional | |

Fig. 3: Experiment environment

|  | Performance improvement of this paper implementation |
|---|---|
| Time domain finite impulse response filter | 4.0 |
| MRI-Q | 7.1 |

Fig. 4: Performance improvement of proposed automatic FPGA offloading method hours to compile one offload pattern.

## 6. Conclusion

As an elementary technology of the environment adaptive software, in this paper, I proposed and evaluated automatic FPGA offloading method for loop statements of software codes. It is often said that FPGA is energy effective than CPU.

The proposed automatic FPGA offloading method is the same as the GPU offloading method [32] until loop statement detection by analyzing the source code. However, it takes a long time to compile to the actual FPGA. In order to cope with the long time compile, the loop statements of offloading candidates are narrowed down before the actual measurement trials are performed. For the detected loop statements, a loop statement having a high arithmetic intensity is extracted using an arithmetic intensity analysis tool. Then, pre-compile is performed to conduct FPGA offloading such as expansion processing for loop statements with high arithmetic intensity. This finds a loop statement with high resource efficiency and high arithmetic intensity. For the narrowed-down loop statements, our method generates OpenCL codes that offload each loop statement or the combination of those loop statements, compiles them on the FPGA, measures the performances and selects high performance OpenCL code as the solution.

## 7. References

[1] A. Putnam, et al., "A reconfigurable fabric for accelerating large-scale datacenter services," ISCA'14, pp.13-24, 2014.

[2] AWS EC2 web site, https://aws.amazon.com/ec2/instance-types/

[3] O. Sefraoui, et al., "OpenStack: toward an open-source solution for cloud computing," International Journal of Computer Applications, Vol.55, 2012.

[4] Y. Yamato, et al., "Fast and Reliable Restoration Method of Virtual Resources on OpenStack," IEEE Transactions on Cloud Computing, Sep. 2015.

[5] Y. Yamato, "Cloud Storage Application Area of HDD-SSD Hybrid Storage, Distributed Storage and HDD Storage," IEEJ Transactions on Electrical and Electronic Engineering, Vol.11, pp.674-675, Sep. 2016.

[6] Y. Yamato, "Use case study of HDD-SSD hybrid storage, distributed storage and

HDD storage on OpenStack," 19th International Database Engineering & Applications Symposium (IDEAS15), pp.228-229, July 2015.

[7] Y. Yamato, "Automatic verification technology of software patches for user virtual environments on IaaS cloud," Journal of Cloud Computing, Springer, 2015, 4:4, Feb. 2015.

[8] Y. Yamato, "Optimum Application Deployment Technology for Heterogeneous IaaS Cloud," Journal of Information Processing, Vol.25, No.1, pp.56-58, Jan. 2017.

[9] Y. Yamato, "Performance-Aware Server Architecture Recommendation and Automatic Performance Verification Technology on IaaS Cloud," Service Oriented Computing and Applications, Springer, Nov. 2016.

[10] Y. Yamato, "Server Selection, Configuration and Reconfiguration Technology for IaaS Cloud with Multiple Server Types," Journal of Network and Systems Management, Springer, Aug. 2017.

[11] Y. Yamato, et al., "Development of Low User Impact and Low Cost Server Migration Technology for Shared Hosting Services," IEICE Transactions on Communications, Vol.J95-B, No.4, pp.547-555, Apr. 2012.

[12] Y. Yamato, et al., "Software Maintenance Evaluation of Agile Software Development Method Based on OpenStack," IEICE Transactions on Information ¥& Systems, Vol.E98-D, No.7, pp.1377-1380, July 2015.

[13] Y. Yamato, "OpenStack Hypervisor, Container and Baremetal Servers Performance Comparison," IEICE Communication Express, Vol.4, No.7, pp.228-232, July 2015.

[14] J. E. Stone, et al., "OpenCL: A parallel programming standard for heterogeneous computing systems," Computing in science & engineering, Vol.12, No.3, pp.66-73, 2010.

[15] J. Sanders and E. Kandrot, "CUDA by example : an introduction to general-purpose GPU programming," Addison-Wesley, 2011

[16] M. Hermann, et al., "Design Principles for Industrie 4.0 Scenarios," Rechnische Universitat Dortmund. 2015.

[17] Y. Yamato, "Experiments of posture estimation on vehicles using wearable acceleration sensors," The 3rd IEEE International Conference on Big Data Security on Cloud (BigDataSecurity 2017), pp.14-17, May 2017.

[18] Y. Yamato, et al., "Predictive Maintenance Platform with Sound Stream Analysis in Edges," Journal of Information Processing, Vol.25, pp.317-320, Apr. 2017.

[19] Tron project web site, http://www.tron.org/

[20] P. C. Evans and M. Annunziata, "Industrial Internet: Pushing the Boundaries of Minds and Machines," Technical report of General Electric (GE), Nov. 2012.

[21] Y. Yamato, "Ubiquitous Service Composition Technology for Ubiquitous

Network Environments," IPSJ Journal, Vol.48, No.2, pp.562-577, Feb. 2007.

[22] H. Sunaga, et al., "Service Delivery Platform Architecture for the Next-Generation Network," ICIN2008, 2008.

[23] Y. Yamato, et al., "Study of Service Processing Agent for Context-Aware Service Coordination," IEEE SCC 2008, pp.275-282, July 2008.

[24] Y. Yamato, et al., "Study and Evaluation of Context-Aware Service Composition and Change-Over Using BPEL Engine and Semantic Web Techniques," IEEE CCNC 2008, pp.863-867, Jan. 2008.

[25] Y. Yamato, et al., "Development of Service Control Server for Web-Telecom Coordination Service," IEEE ICWS 2008, pp.600-607, Sep. 2008.

[26] Y. Yokohata, et al., "Service Composition Architecture for Programmability and Flexibility in Ubiquitous Communication Networks," IEEE International Symposium on Applications and the Internet Workshops (SAINTW'06), 2006.

[27] H. Sunaga, et al., "Ubiquitous Life Creation through Service Composition Technologies," World Telecommunications Congress 2006 (WTC2006), May 2006.

[28] Y. Nakano, et al., "Effective Web-Service Creation Mechanism for Ubiquitous Service Oriented Architecture," The 8th IEEE International Conference on E-Commerce Technology and the 3rd IEEE International Conference on Enterprise Computing, E-Commerce, and E-Services (CEC/EEE 2006), pp.85, June 2006.

[29] Y. Yamato, et al., "Study of Service Composition Engine Implemented on Cellular Phone," Information technology letters, Vol.4, pp.269-271, Aug. 2005.

[30] Y. Yokohata, et al., "Context-Aware Content-Provision Service for Shopping Malls Based on Ubiquitous Service-Oriented Network Framework and Authentication and Access Control Agent Framework," IEEE CCNC 2006, pp.1330-1331, 2006.

[31] J. Gosling, et al., "The Java language specification, 3rd edition," Addison-Wesley, 2005.

[32] Y. Yamato, et al., "Automatic GPU Offloading Technology for Open IoT Environment," IEEE Internet of Things Journal, Sep. 2018.

[33] Y. Yamato, "Study of parallel processing area extraction and data transfer number reduction for automatic GPU offloading of IoT applications," Journal of Intelligent Information Systems, Springer, DOI:10.1007/s10844-019-00575-8, 2019.

[34] Altera SDK web site, https://www.altera.com/products/design-software/ embedded-software-developers/opencl/documentation.html

[35] Xilinx, http://japan.xilinx.com/html_docs/xilinx2017_4/sdaccel_doc/lyx1504034296578

[36] K. Shirahata, et al., "Hybrid Map Task Scheduling for GPU-Based Heterogeneous Clusters,"IEEE CloudCom, 2010.

[37] S. Wienke, et al., "OpenACC-first experiences with real-world applications,"

Euro-Par Parallel Processing, 2012.

[38] M. Wolfe, "Implementing the PGI accelerator model," ACM the 3rd Workshop on General-Purpose Computation on Graphics Processing Units, pp.43-50, Mar. 2010.

[39] K. Ishizaki, "Transparent GPU exploitation for Java," CANDAR 2016, Nov. 2016.

[40] E. Su, et al., "Compiler support of the workqueuing execution model for Intel SMP architectures," In Fourth European Workshop on OpenMP, Sep. 2002.

[41] Jenkins web site, https://jenkins.io/

[42] Selenium web site, https://www.seleniumhq.org/

[43] Clang website, http://llvm.org/

[44] OpenCV web site, http://opencv.org/

[45] imageJ web site, https://imagej.nih.gov/ij/docs/concepts.html

[46] J. H. Holland, "Genetic algorithms," Scientific american, Vol.267, No.1, pp.66-73, 1992.

[47] gcov website, http://gcc.gnu.org/onlinedocs/gcc/Gcov.html

[48] Time Domain Finite Impulse Response Filter web site, http://www.omgwiki.org/hpec/files/hpec-challenge/tdfir.html

[49] MRI-Q website, http://impact.crhc.illinois.edu/parboil/parboil.aspx